\pdfoutput=1
\documentclass[twocolumn,aps,prd]{revtex4}
\usepackage{graphicx,amssymb}
\usepackage[utf8]{inputenc} 
\usepackage{hyperref}
\usepackage{amsmath}

\begin{document}
\def\half{{1\over2}}
\def\shalf{\textstyle{{1\over2}}}

\newcommand\lsim{\mathrel{\rlap{\lower4pt\hbox{\hskip1pt$\sim$}}
    \raise1pt\hbox{$<$}}}
\newcommand\gsim{\mathrel{\rlap{\lower4pt\hbox{\hskip1pt$\sim$}}
    \raise1pt\hbox{$>$}}}

\title{Big-bang nucleosynthesis and cosmic microwave background constraints on non-minimally coupled theories of gravity}

\author{R.P.L. Azevedo}
\email[Electronic address: ]{rplazevedo@fc.up.pt}
\affiliation{Instituto de Astrof\'{\i}sica e Ci\^encias do Espa{\c c}o, Universidade do Porto, CAUP, Rua das Estrelas, PT4150-762 Porto, Portugal}
\affiliation{Centro de Astrof\'{\i}sica da Universidade do Porto, Rua das Estrelas, PT4150-762 Porto, Portugal}
\affiliation{Departamento de F\'{\i}sica e Astronomia, Faculdade de Ci\^encias, Universidade do Porto, Rua do Campo Alegre 687, PT4169-007 Porto, Portugal}

\author{P.P. Avelino}
\email[Electronic address: ]{pedro.avelino@astro.up.pt}
\affiliation{Instituto de Astrof\'{\i}sica e Ci\^encias do Espa{\c c}o, Universidade do Porto, CAUP, Rua das Estrelas, PT4150-762 Porto, Portugal}
\affiliation{Centro de Astrof\'{\i}sica da Universidade do Porto, Rua das Estrelas, PT4150-762 Porto, Portugal}
\affiliation{Departamento de F\'{\i}sica e Astronomia, Faculdade de Ci\^encias, Universidade do Porto, Rua do Campo Alegre 687, PT4169-007 Porto, Portugal}

\date{\today}
\begin{abstract}

In this paper we show that the baryon-to-photon ratio $\eta$ is in general not conserved before decoupling in non-minimally coupled theories of gravity. We use big-bang nucleosynthesis and cosmic microwave background limits on the baryon-to-photon ratio $\eta$ to derive new constraints on modified gravity theories with a universal non-minimal coupling between matter and curvature, showing that they rule out a specific class of models previously considered in the literature as a substitute for the dark matter. We also compare these new constraints with the ones obtained from the COBE-FIRAS limits on CMB spectral distortions, highlighting the complementarity between them.

\end{abstract}
\maketitle

\section{Introduction}
\label{sec:intr}

The origin of the accelerated expansion of the Universe and of the non-trivial dynamics of galactic disks and clusters are two of the greatest enigmas facing modern cosmology \cite{Bertone2005,Carroll2001}. The standard solution to these fundamental questions   relies on the assumption that the cosmological dynamics are well described by General Relativity (GR) and that the Universe is filled not only with baryonic matter and radiation, but also with cold dark matter, playing a crucial role in the observed galactic dynamics, and dark energy, responsible for the accelerated expansion of the Universe. Alternatively, one can assume that GR is incomplete, and that there is a more accurate theory at work on cosmological scales which  eliminates the need for dark energy (or even dark matter --- see, however, \cite{Clowe2004,Markevitch2004}). Many extensions of GR have been proposed in the literature, such as theories with additional fields, theories with more complex geometric terms such as $f(R)$ and $f(R, R_{\mu\nu}, R_{\mu\nu\alpha\beta})$, and theories featuring a non-minimal coupling (NMC) between geometry and matter, such as $f(R,\mathcal{L}_m)$ theories \cite{Nojiri2004,Allemandi2005,Bertolami2007,Sotiriou2008,Harko2010,Harko2011}.

NMC theories can introduce significant modifications to the gravitational dynamics with a particular impact on  cosmology. A fundamental property of these theories is that the energy-momentum tensor is not covariantly conserved, with the Lagrangian of the matter fields entering explicitly in the equations of motion. This makes the use of the correct form of the Lagrangian of fundamental importance in this context. In previous work $\mathcal{L}_m=-\rho$ or $\mathcal{L}_m=p$ have been suggested as the Lagrangian of a perfect fluid \cite{Nesseris2009,Bertolami2012,Azizi2014,Ribeiro2014,Silva2018}. However, it has recently been shown that the correct Lagrangian for a fluid composed of solitonic particles of fixed rest mass and structure is given by the trace of the energy-momentum tensor of the fluid $\mathcal{L}_m = T = 3p-\rho$ \cite{Avelino2018a,Avelino2018}. This is expected to be a good approximation in the case of baryonic matter, dark matter and photons (the zero rest mass limt being considered in the case of photons), but it does not apply to dark energy or to any fluid with an equation of state parameter outside the interval $0\leq w \leq 1/3$. 

In Ref. \cite{Avelino2018} a new source of spectral distortions ($n$-type spectral distortions) of the CMB power spectrum has been studied in the context of NMC theories of gravity and the measurement of the black-body spectrum by the Far-Infrared Absolute Spectrophotometer (FIRAS), on board the Cosmic Background Explorer (COBE), was used to put stringent limits on NMC gravity. In the present work Big-Bang Nucleosynthesis (BBN) and Cosmic Microwave Background (CMB) constraints are used to probe NMC gravity at even higher redshifts. The structure of the paper is as follows. In Section \ref{sec:model} we present the NMC gravity action, derive the cosmological equations of motion and discuss the dependence of the evolution of the energy density on the equation of state parameter of each fluid component. In Section \ref{sec:BBN} we discuss the changes to BBN arising in the context of NMC theories of gravity, and obtain the corresponding constraints using BBN and CMB limits on the baryon-to-photon ratio $\eta$. We also determine the constraints on a specific sub-class of models previously considered in the literature as a substitute for dark matter, which are then compared with those obtained from the COBE-FIRAS limits on CMB spectral distortions. The conclusions of this work are presented in Section \ref{sec:conc}.

Throughout this paper we use fundamental units such that $c=\hbar=k_B=1$. Here $c$ is the value of the speed of light in vacuum, $\hbar=h/(2\pi)$ where $h$ is the Planck constant, and $k_B$ is the Boltzmann constant. We adopt the metric signature $(-,+,+,+)$, and the Einstein summation convention will be used as usual.


\section{Non-minimally coupled gravity}
\label{sec:model}

Consider the action
\begin{equation}
\label{eq:action}
S=\int \sqrt{-g} \left[\kappa f_1(R) + f_2(R)\mathcal{L}_m\right]\,,
\end{equation}
where $\kappa=(16\pi G)^{-1}$, $G$ is Newton's gravitational constant, $g$ is the determinant of the metric $g_{\mu\nu}$, $\mathcal{L}_m$ is the Lagrangian of the matter fields, and $f_1(R)$ and $f_2(R)$ are generic functions of the Ricci scalar $R$. GR is recovered if $f_1(R)=R$ and $f_2(R)=1$. Extremizing the action with respect to the metric one obtains the equation of motion of the gravitational field
\begin{equation}
\label{eq:eqmotion}
F G_{\mu\nu}=\half f_2 T_{\mu\nu}+\Delta_{\mu\nu}F+\half\kappa f_1 g_{\mu\nu}-\half RFg_{\mu\nu}\,,
\end{equation}
where $G_{\mu\nu}=R_{\mu\nu}-\shalf g_{\mu\nu} R $ is the Einstein tensor, $R_{\mu\nu}$ is the Ricci tensor, $\Delta_{\mu \nu} \equiv \nabla_\mu \nabla_\nu - g_{\mu \nu} \Box$, $\Box \equiv \nabla^\mu \nabla_\mu$,  
\begin{equation}
\label{eq:F}
F=\kappa f'_1(R)+f'_2(R)\mathcal{L}_m\,,
\end{equation}
a prime denotes a derivative with respect to the Ricci scalar, and the energy-momentum tensor has the usual form
\begin{equation}
\label{eq:energymom}
T_{\mu\nu}=-{2\over \sqrt{-g}}{\delta(\sqrt{-g}\mathcal{L}_m)\over \delta g^{\mu\nu}}\,.
\end{equation}

A crucial feature of these theories is that the energy-momentum tensor is no longer covariantly conserved: in fact, applying the Bianchi identities to the equations of motion leads to
\begin{equation}
\label{eq:noncons}
\nabla^\mu T_{\mu\nu}={f'_2\over f_2}(g_{\mu\nu}\mathcal{L}_m-T_{\mu\nu})\nabla^\mu R\, ,
\end{equation}
so  though we consider that the NMC does not significantly affect the structure of particles, the explicit presence of the matter Lagrangian in Eqs. \eqref{eq:eqmotion} and \eqref{eq:noncons} is of crucial importance, as it directly affects particle motion \cite{Bertolami2008,Avelino2018}.

In order to study the evolution of a flat homogeneous and isotropic universe one has to consider the flat Friedmann-Robertson-Walker metric with line element
\begin{equation}
\label{eq:metric}
ds^2=-dt^2+a^2(t)\left[dx^2 + dy^2 +dz^2\right]\,,
\end{equation}
where $a(t)$ is the scale factor, $t$ is the cosmic time, and $x$, $y$, and $z$ are Cartesian comoving coordinates. The energy content of the Universe (except for dark energy) will be assumed to be described by a perfect-fluid with energy-momentum tensor
\begin{equation}
\label{eq:pfemt}
T^{\mu\nu}=(\rho+p)u^\mu u^\nu + p g^{\mu\nu}\,,
\end{equation}
where $\rho$, $p$ and $u^\mu$ are, respectively, the energy density, pressure and four-velocity of the fluid. This form of the energy-momentum tensor is associated to the Lagrangian
\begin{equation}
\label{eq:lagrangian}
\mathcal{L}_m=T=3p-\rho\,,
\end{equation}
where $T$ is the trace of the energy-momentum tensor, if one considers that the fluid can be described, at a microscopic level, by localized concentrations of energy  (solitonic particles) \cite{Avelino2018a,Avelino2018}. While the specific structure of the particles will not be relevant for the present study, any change to their structure and mass will be assumed to be negligible \cite{Uzan2011,Copeland2007}.

A key feature of this Lagrangian is that some fluids may follow the usual conservation law, while others do not. In fact, taking the time component of Eq. \eqref{eq:noncons} and using Eq. \eqref{eq:lagrangian}, one obtains
\begin{equation}
\label{eq:densityevolution}
\dot{\rho}=-3\rho\left[H(1+w)+w{f'_2\over f_2}\dot{R}\right]\,,
\end{equation} 
where $H\equiv \dot{a}/a$ is the Hubble parameter, a dot represents a derivative with respect to the cosmic time, and $w=p/\rho$ is the equation of state parameter. For a single fluid $i$, Eq. \eqref{eq:densityevolution} can be directly integrated to give
\begin{equation}
\label{eq:densityevo}
\rho_i=\rho_{i,0} a^{-3(1+w_i)} f_2^{-3w_i}\,,
\end{equation}
where $\rho_{i,0}$ is the energy density at the present time, when $a(t)=a_0=1$. It is then immediate to see that in the case of dust (with $w=0$) the usual conservation law $\rho \propto a^{-3}$ holds, while in the case of photons (with $w=1/3$) the NMC generally leads to a significant change to the evolution of the photon energy density ($\rho \propto a^{-4}f_2^{-1}$ instead of $\rho \propto a^{-4}$). The relative change to the conservation laws of photons and baryons can then be used to derive strong constraints on the form of $f_2$.

The $tt$ component of Eq. \eqref{eq:eqmotion} yields the modified Friedmann equation
\begin{equation}
\label{eq:modfried}
H^2={1\over 3F}\left[\half \left(FR- \kappa f_1+ f_2 \rho\right)-3H\dot{F}\right]\,,
\end{equation}
while the $rr$ component constitutes the modified Raychaudhury equation
\begin{equation}
\label{eq:modray}
H^2=-{1\over 2F}\left[{1\over 3}FR-\kappa f_1-f_2 w\rho -2\ddot{F}-4H\dot{F}\right]\,,
\end{equation}
and the trace reads
\begin{equation}
\label{eq:tracemotion}
FR=2\kappa f_1+\half f_2 (3w-1)\rho +9H\dot{F}+3\ddot{F}\,,
\end{equation}
where we have taken into account that $R=6(\dot{H}+2H^2)$.

\section{Big-bang nucleosynthesis}\label{sec:BBN}

One of the great successes of modern cosmology is the prediction of the abundances of light elements formed in the early Universe. As the the soup of electrons, positrons, photons, neutrinos and nucleons cooled down due to cosmological expansion, the weak interaction rates eventually dropped below the expansion rate $H$, with neutrinos departing from thermodynamic equilibrium with the remaining plasma. In the context of the present work, the most relevant consequence of this phenomenon is the breaking of the neutron-proton chemical equilibrium at $T_D\sim 0.7 \text{ MeV}$, which leads to the freeze out of the neutron-proton number density ratio at $n_n/n_p=\exp(-\Delta m/T_D) \sim 1/7$, where $\Delta m = 1.29 \text{ MeV}$ is the neutron proton mass difference (the ratio $n_n/n_p$ is then  slightly reduced by subsequent neutrons decays). Soon after, at $T_N\sim 100 \text{ keV}$, the extremely high photon energy density has been diluted enough to allow for the formation of the first stable $^2$H deuterium nuclei.

Once $^2$H starts forming, an entire nuclear process network is set in motion, leading to the production of light-element isotopes and leaving all the decayed neutrons bound into them, the vast majority in $^4$He nuclei. Primordial nucleosynthesis may be described by the evolution of a set of differential equations, namely the Friedmann equation, the evolution of baryon and entropy densities, and the Boltzmann equations describing the evolution of the average density of each nuclide and neutrino species. As one could expect, even taking into account experimental values for the reaction cross-sections instead of theoretical derivations from particle physics, the accurate computation of element abundances cannot be done without resorting to numerical algorithms \cite{Wagoner1973,Kawano1992,Smith:1992yy,Pisanti2008,Consiglio2017}. The prediction of these quantities in the context of NMC theories is beyond the scope of the present paper, but it is worthy of note that these codes require one particular parameter to be set a priori: the baryon-to-photon ratio $\eta$.

While in GR the baryon-to-photon ratio is fixed around nucleosynthesis, the same does not in general happen in the context of NMC theories. To show this, recall that the evolution of the density of photons and baryons (which are always non-relativistic from the primordial nucleosynthesis epoch up to the present era) is given by Eq. \eqref{eq:densityevo} as
\begin{equation}
\label{eq:matradevo}
\rho_\gamma = \rho_{\gamma,0} a^{-4}f_2^{-1}\,, \qquad \rho_B =\rho_{B,0} a^{-3}\,.
\end{equation}
While the baryon number in a fixed comoving volume is conserved ($n_B\propto a^{-3}$, where $n_B$ is the baryon number density), before recombination photons are in thermal equilibrium, so the photon number density is directly related to the temperature by
\begin{equation}
\label{eq:photntemp}
n_\gamma={2\zeta(3)\over \pi^2}T^3\,.
\end{equation}
Since the photon energy density also relates to the temperature as
\begin{equation}
\label{eq:photenergdens}
\rho_\gamma={\pi^2\over 15}T^4\,,
\end{equation}
combining Eqs. \eqref{eq:matradevo}, \eqref{eq:photntemp} and \eqref{eq:photenergdens} one obtains that that the baryon-to-photon ratio $\eta$ between BBN (at a redshift $z_{BBN} \sim 10^9$) and photon decoupling (at a redshift $z_{CMB} \sim 10^3$) evolves as
\begin{equation}
\label{eq:eta}
\eta\equiv {n_B\over n_\gamma}\propto f_2^{3/4}\,,
\end{equation}
as opposed to the GR result, $\eta=\text{const}$. After photon decoupling the baryon-to-photon ratio is conserved in NMC theories, with the energy of individual photons evolving as $E_\gamma \propto (a f_2)^{-1}$ \cite{Avelino2018} (as opposed to the standard $E_\gamma \propto a^{-1}$ result).

We will assume that the modifications to the dynamics of the universe with respect to GR are small, in particular to the evolution of $R$ and $H$ with the redshift $z$. This can be seen as a ``best-case'' scenario for the theory, as any significant changes to $R(z)$ and $H(z)$ are expected to worsen the compatibility between the model's predictions and observational data. Hence, in the following we shall assume that 
\begin{eqnarray}
R &=& 3 H_0^2 \left[ \Omega_{m,0} (1+z)^{3} + 4 \Omega_{\Lambda,0} \right] \nonumber  \\
&\sim&  3 H_0^2 \Omega_{m,0} (1+z)^{3} \propto (1+z)^{3}\,,
\end{eqnarray}
where $\Omega_{m,0} \equiv (\rho_{m,0})/(6 \kappa H_0^2)$ and $\Omega_{\Lambda,0} \equiv (\rho_{\Lambda,0})/(6 \kappa H_0^2)$ are the matter and dark energy density parameters (here dark energy is modelled as a cosmological constant $\Lambda$), and the approximation is valid all times, except very close to the present time.
Let us define
\begin{eqnarray}
\Delta f_2^ {i \to f} &\equiv& \left|f_2(z_f) - f_2(z_i)\right| \,, \\
\frac{\Delta \eta}{\eta}^{i \to f} &\equiv& \frac{\left|\eta(z_f) - \eta(z_i)\right|}{\eta(z_i)} \,,
\end{eqnarray}
which shall be both assumed to be much smaller than unity. Here, $z_i$ and $z_f$ are given initial and final redshifts with $z_i > z_f$. Consider a power law model for $f_2$ defined by
\begin{equation}
\label{eq:model}
f_1(R) \sim R\,, \qquad f_2(R) \propto R^n\,,
\end{equation}
in the redshift range $[z_f,z_i]$, where $n$ is a real number with $|n| \ll 1$ and $f_2 \sim 1$ at all times. In this case
\begin{equation}
\label{eq:f2var}
\Delta f_2^ {i \to f}  \sim   \left| \left(\frac{R(z_f)}{R(z_i)}\right)^n -1\right| 
 \sim  3 \left| n \right| \ln \left( \frac{1+z_f}{1+z_i}\right)\,.
\end{equation}
Eq. \eqref{eq:f2var} then implies that
\begin{equation}
\left| n \right| \lsim \frac49 \frac{\Delta \eta}{\eta}^{i \to f} \left[ \ln \left( \frac{1+z_f}{1+z_i}\right) \right]^{-1}\,,
\label{eq:nconstraint}
\end{equation}
assuming a small relative variation of $\eta$ satisfying Eq. \eqref{eq:eta}
\begin{equation}
\label{eq:etavar}
\Delta f_2^ {i \to f}  \sim \frac43 \frac{\Delta \eta}{\eta}^{i \to f}\,.
\end{equation}

There are two main ways of estimating the value of $\eta$ at different stages of cosmological evolution. On one hand, one may combine the observational constraints on the light element abundances with numerical simulations of primordial BBN nucleosynthesis to infer the allowed range of $\eta$. This is the method used in \cite{Iocco2009}, among others, leading to
\begin{equation}
\label{eq:etabbn}
\eta_{BBN}=(5.7\pm0.6)\times 10^{-10}
\end{equation}
at 95\% Confidence Level (CL) just after nucleosynthesis (at a redshift $z_{BBN} \sim10^9$). More recently, an updated version of the program {\fontfamily{qcr}\selectfont PArthENoPE} ({\fontfamily{qcr}\selectfont PArthENoPE} 2.0), which computes the abundances of light elements produced during Big Bang Nucleosynthesis, was used to obtain new limits on the baryon-to-photon ratio, at $2\sigma$ \cite{Consiglio2017}
\begin{equation}
\label{eq:etabbn2}
\eta_{BBN}=(6.23^{+0.24}_{-0.28})\times 10^{-10}\,.
\end{equation}

There is actually some variation of $\eta$ during nucleosynthesis due to the entropy transfer to photons associated to the $e^\pm$ annihilation. The ratio between the values of $\eta$, respectively, at the beginning ($T\simeq10$ MeV) and at the end of BBN is given approximately by a factor of $2.73$ \cite{Serpico2004}. Although the NMC will lead to further changes on the value of $\eta$ during BBN, we will not consider this effect since it will be subdominant for $|n| \ll 1$. We will therefore use the above standard values obtained for $\eta_{BBN}$ immediately after nucleosynthesis to constrain NMC gravity.

The neutron-to-photon ratio also affects the acoustic peaks observed in the CMB, generated at a redshift $z_{CMB} \sim10^3$. The full-mission Planck analysis \cite{Ade2016} constrains the baryon density $\omega_B = \Omega_B (H_0/[100 \text{ km s}^{-1}\text{ Mpc}^{-1}] )$ from baryon acoustic oscillations, at 95\% CL,
\begin{equation}
\label{eq:omegacmb}
\omega_B=0.02229^{+0.00029}_{-0.00027}\, .
\end{equation}
This quantity is related to the baryon-to-photon ratio via $\eta = 273.7\times10^{-10} \omega_B$, leading to
\begin{equation}
\label{eq:etacmb}
\eta_{CMB}= 6.101^{+0.079}_{-0.074}\times 10^{-10}\,.
\end{equation}
Here, we implicitly assume that no significant change to $\eta$ occurs after $z_{CMB} \sim10^3$, as shown in Ref. \cite{Avelino2018} (we will comeback to this point further on).

Taking these results into consideration, we shall determine conservative constraints on $n$ using the maximum allowed variation of $\eta$ from $z_{BBN} \sim10^9$ to $z_{CMB} \sim10^3$, using the appropriate lower and upper limits given by Eqs. \eqref{eq:etabbn}, \eqref{eq:etabbn2} and \eqref{eq:etacmb}. Combining Eqs. \eqref{eq:eta} and \eqref{eq:model} to obtain
\begin{equation}
\label{eq:etamodel}
\eta\propto R^{3n/4}\,,
\end{equation}
it is easy to see that the sign of $n$ will affect whether $\eta$ is decreasing or increasing throughout the history of the universe, and thus, since $R$ monotonically decreases towards the future, a positive (negative) $n$ will imply a decreasing (increasing) $\eta$. This being the case, for the allowed range in Eq. \eqref{eq:etabbn}, we have for positive $n$
\begin{equation}
\label{eq:npos}
\frac{\Delta \eta}{\eta} = \frac{\left|(6.101 - 0.074) - (5.7 + 0.6)\right|}{5.7 + 0.6} \simeq 0.04 \,,
\end{equation}
and for negative $n$
\begin{equation}
\label{eq:nneg}
\frac{\Delta \eta}{\eta} = \frac{\left|(6.101 + 0.079) - (5.7 - 0.6)\right|}{5.7 - 0.6} \simeq 0.21 \,,
\end{equation}
Therefore we find
\begin{equation}
\label{eq:nconstr1}
-0.007<n<0.002\, ,
\end{equation}
and using the limits given in Eq. \eqref{eq:etabbn2} \cite{Consiglio2017},
\begin{equation}
\label{eq:nconstr2}
-0.002<n<0.003\, .
\end{equation}

In \cite{Avelino2018} the NMC has been shown to lead to $n$-type spectral distortions, affecting the normalization of the spectral energy density. In this work the variation of the function $f_2$ has been constrained to a few parts into $10^5$ form the redshift of photon decoupling ($z_{CMB} \sim10^3$) up to the present time $z_0 = 0$ (a similar constraint applies to the variation of $\eta$). Taking into account that
\begin{equation}
\left| n \right| \lsim \frac{\Delta f_2}{3}^{CMB \to 0} \left[ \ln \left(1+z_{CMB}\right) \right]^{-1}\,.
\label{eq:nconstraint3}
\end{equation}
it is simple to show that this translates into $|n|\lesssim {\rm few} \times 10^{-6}$, which is roughly $3$ orders of magnitude stronger than the constraint coming from the baryon-to-photon ratio. Still, this constraint and those given by Eqs. \eqref{eq:nconstr1} and \eqref{eq:nconstr2} are associated to cosmological observations which probe different epochs and, as such, can be considered complementary: while the former limits an effective value of the power law index $n$ in the redshift range $[0,10^3]$, the later is sensitive to its value at higher redshifts in the range $[10^3,10^9]$.

Furthermore, NMC theories with a power-law coupling $f_2(R)$ have been considered as a substitute for dark matter in previous works \cite{Bertolami2012,Silva2018}. There it has been shown that $n$ would have to be in the range $-1\leq n \leq -1/7$ in order to explain the observed galactic rotation curves. However, such values of $n$ are totally excluded by the present study.

\section{Conclusions}\label{sec:conc}

In this work we have shown that non-minimal coupling to photons may cause the baryon-to-photon ratio, which is constrained both using primordial element abundances and the CMB acoustic peaks, to change between the epochs of primordial nucleosynthesis and photon decoupling. This has been used to derive new constraints on theories that feature an $f(R)$-inspired non-minimal coupling to matter. For a power-law coupling, previously considered in the literature as a substitute for the dark matter, we have shown that NMC gravity is excluded as a possible explanation to the observed galactic rotation curves, a result that may only be relaxed if gravity does not couple non-minimally to photons.

\begin{acknowledgments}
R.P.L.A. was supported by the Funda{\c c}\~ao para a Ci\^encia e Tecnologia (FCT, Portugal) grant SFRH/BD/132546/2017. Funding of this work has also been provided by the FCT grant UID/FIS/04434/2013. This paper benefited from the participation of the authors on the COST action CA15117 (CANTATA), supported by COST (European Cooperation in Science and Technology).
\end{acknowledgments}

\bibliography{BBN_constraints_NMC}

\end{document}